# Localized Polymerization Using Single Photon Photoinitiators in Two-photon process for Fabricating Subwavelength Structures

*Govind Ummethala, Arun Jaiswal, Raghvendra P Chaudhary, Suyog Hawal and Shobha Shukla\**

Nanostructures Engineering and Modeling Laboratory, Department of Metallurgical Engineering and Materials Science, Indian Institute of Technology Bombay, Mumbai, MH, India 400076.

Abstract

Localized polymerization in subwavelength volumes using two photon dyes has now become a well-established method for fabrication of subwavelength structures. Unfortunately, the two photon absorption dyes used in such process are not only expensive but also proprietary. LTPO-L is an inexpensive, easily available single photon photoinitiator and has been used extensively for single photon absorption of *UV* light for polymerization. These polymerization volumes however are not localized and extend to micron size resolution having limited applications. We have exploited high quantum yield of radicals of LTPO-Lfor absorption of two photons to achieve localized polymerization in subwavelength volumes, much below the diffraction limit. Critical concentration (10wt%) of LTPO-Lin acrylate (Sartomer) was found optimal to achieve subwavelength localized polymerization and has been demonstrated by fabricating 2D/3D complex nanostructures and functional devices such as variable polymeric gratings with nanoscaled subwavelength resolution. Systematic studies on influence of LTPO-L concentration on two photon polymerization of commercially available photopolymer (Sartomer) show that resolution of the fabricated structures critically depend on loading of LPTO-L. This is expected to unleash the true potential of two photon polymerization for fabrication of complex polymeric nanodevices at a larger scale.

\*corresponding author - Tel. no. +91-22-25767615. Email-sshukla@iitb.ac.in

# 1. Introduction

The developments in the field of polymers for over a century has enabled their applications ranging from high tech devices to simple polyethylene carry bag used by a common man. Polymerization involves reaction of monomer units to form large networks, involving external stimulants such as catalyst, plasma, photons and the like. Photo-polymerization is one such technique for synthesizing polymers and in general is a chain growth polymerization. In this process polymerization is initiated by absorption of photons either directly by the reactant monomer or by energy transfer through absorption of energy by a photosensitizer. Two photon polymerization (TPP) involves absorption of two photons simultaneously through a virtual state for excitation of a molecule. It has emerged as an extremely powerful tool of triggering chemical and physical processes with high spatial resolutionfor fabrication of 2D/3D nanostructures via a single step, which is scalable, comparatively inexpensive and a vacuum less approach[1–7]. The quadratic dependence of the two photon process confines the absorption of photons at the focal point within a volume of the order $\lambda^3$ enabling sub-wavelength structures to be fabricated. This process has been of particular interest to fabricate complex micro/nanostructures for optical circuitry, optical data storage,biology and micro-fabrication technologiesand photonic crystals [4,8,9,10,11].

An essential requirement for a materialto be used as a two photon photoinitiator is its optical transparency to wavelength of ultra-short pulse laser, which typically liesin the region between 600 and 800 nm. This ensuresminimization of linear absorption by the photointiator/chromophore which may occur during TPP.The two photon absorption (TPA) cross section is generallyused as a metric for comparison of two photon absorption activities of different dye/photoinitiators[12]. Typically two photon absorption dyes are used for this purpose.A highly sensitive and efficient dye can lead to very low threshold and short exposure time, which would lower the volume where radicals are initially generated and decrease the amount of radicals formed and diffusion of the same[13]. Variousstrategies used to design a good TPP initiatorinclude a chromophore group with a large $\delta_{TPA}$, such as a D-π-D structure, chemical functionality that has a high efficiency of initiation and mechanisms by which excitation of the chromophore leads to activation of an electron-transfer process[14]. Efforts have been made to synthesize several linear and cyclic benylidene ketone based TPP initiators containing double bonds and dialkylamino groups in a single step using aldol condensation reactions. Quantum chemical calculations and experimental tests conducted to determine the structural – activity relationships proved that the size of the central ring of the TPP initiators significantly impacted the TPP initiation efficiency[15]. The lateral spatial resolution LSR in two-photon induced polymerization was improved to 80 nm by using an anthracene derivative 9,10-bis-pentyloxy-2,7-bis2-4-dimethylamino-phenyl-vinylanthracene BPDPA as a highly sensitive and efficient photo-initiator. Though a lot of molecules with very high TPA cross sections have been developed, their toxicity and commercial unavailability limits the applicability[16].

*corresponding author - Tel. no. +91-22-25767615. Email-sshukla@iitb.ac.in

A class of commercially available radical, single photon photoinitiators such as Irgacure 369, Lucirin TPOL and WLPI have been explored for TPP process[17–20]. These have howeverresulted in increased line-widths as their two photon absorption coefficient is small(~1GM) in the IR region (800nm-1000nm) compared to IR absorbing dyes ($10^3$-$10^5$GM)[13].In spite of this drawbackLTPO-L an acylphosphine oxide radical photoinitiator is best in this class due to several advantages for two photon induced polymerization. Unlike most radical photoinitiators LTPO-L is a solid with good solubility and can be mixed with most resins. Although the two photon absorption cross section is small (< 1.2GM) it has been shown that structures with excellent integrity and definition can be fabricated at relatively low laser powers due to the high radical quantum yield of 0.99. This high polymerization efficiency enables for excellent initiation in two photon lithography over broad spectral ranges[21]. The use of LTPO-L as photoinitiator for TPP by Baldacchini et.al in their Sartomer resin formulation at very low concentrations of (3% weight) gave polymerized features with smallest dimensions of 5 μm[22]. A few more groups have used such formulations for fabrication of polymeric micro-cantilevers, optically active microstructures, gold doped structures and 3D cell migration studies obtaining feature sizes in the order of few microns[23–26]. Very recently branched hollow fiber structures were fabricated with predefined circular pores using Ormocer which is an organic – inorganic hybrid material pre-loaded with Lucirin TPO-L. Well defined micro structureswere obtained by using average laser power of 105 mW and writing speed of 5 mm/s [27]. However, most of these processes resulted in larger linewidths and were inappropriate for controlled polymerization in smaller focal volume to fabricate finer sub wavelength structures.

In this manuscript we exploit the high quantum yield of radicals of single photon photoinitiator (LTPO-L)for use as an alternative to two photon dye for fabrication of 2D and 3D structures with sub-wavelength resolution by using a critical loading of such inexpensive single photon photoinitiators.The high quantum yield of the radicals on exposure of LTPO-L to IR radiation facilitates initiation of the polymerization reactions and makes up for the large two photon absorption coefficient of inefficient two photon dyes.We have further systematically studied the effect of loading concentrations of LTPO-L on the spatial resolution of the fabricated structures and laser power required for lithography. The structures written were extensively characterized using scanning electron microscopy in order to determine the linewidths. The optimized parameters were used to successfully write complex 2D &3D sturdy structures with sub-wavelength resolution.

*corresponding author - Tel. no. +91-22-25767615. Email-sshukla@iitb.ac.in

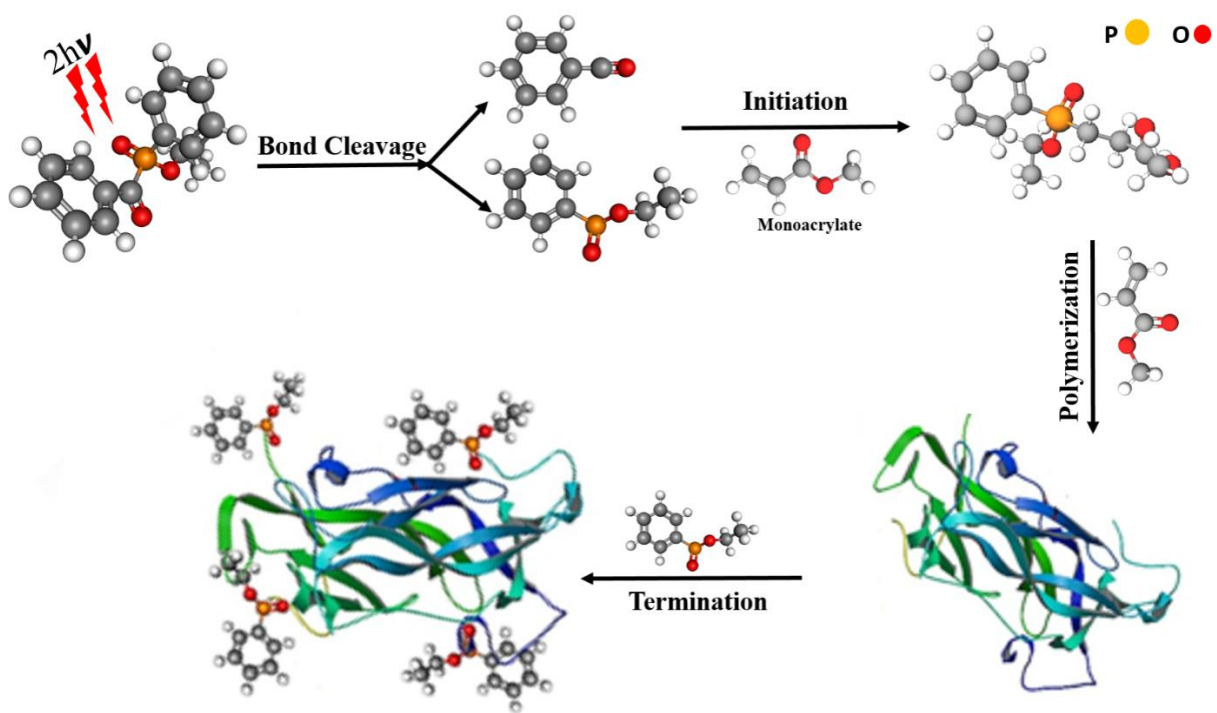

Fig1: Schematic of the free radical chain polymerization reaction which is understood to occur during fabrication. The oxygen atoms are represented by red dots, carbon atoms by grey, hydrogen atoms by white and phosphorous atoms by orange dots. Two colored ribbons represents two polymers used namely SR368 and SR499.

Schematic of free radical chain polymerization reaction is shown in *Fig1* wherein the photoinitiator absorbs incident photons and generates active radicals by chemical decomposition. This usually happens in the triplet state and follows the Norrish type I mechanism. LTPO-L undergoes bond cleavage and results in the formation of two radicals capable of initiating the polymerization reaction. However, due to the high reactivity of the phosphorous based radical, it is anticipated that larger concentration of polymer contained ethyl phenyl phosphinate at the end of the chain. The radicals generated combine with the monomeric units creating reactive centers. The chain polymerization reaction propagates as the reactive centers further combine with other monomeric units, resulting in increase in molecular weight of the polymer until the point of termination.Termination occurs when radical species similar to those responsible for initiation combine with reactive centers, thus turning the growing chain into a dead polymer by undergoing electron transfer.

*corresponding author - Tel. no. +91-22-25767615. Email-sshukla@iitb.ac.in

2. Experimental

The resin used in this study consists of a mixture of two tri-acrylate monomers (Sartomer) and LTPO-L as a photo initiator (BASF). Tris (2–hydroxyethyl) isocyanurate (SR368) increases the hardness of the microstructure whereas the ethoxylated (6) trimethylolpropane tracrylate (SR499) reduces structural shrinkage upon polymerization. All the chemicals were used as received without any further processing.The solutions were prepared by mixing equal amounts of SR 368 and SR 499 to which LTPO-L was added. Solutions with 1 wt%, 3wt%, 5 wt%, 10 wt% and 20 wt% loading of the photo initiator were prepared. This mixture was then uniformly spin coated on a microscopic glass slide to obtain a uniform thin film. Prior to spin coating the slides were treated with 3-aminopropyl triethoxysilane to enhance the adhesion of the final structure to the substrate. The acrylate coated cover slip was positioned on the XYZ piezo stage of an inverted optical microscope (Olympus-IX81). The samples were irradiated with a Ti- Sapphire femto second laser of 800nm wavelength and 140fs pulse width at a repetition rate of 80MHz (Coherent Chameleon, Ultra I). A 100x objective with NA of 0.9 was employed to focus the laser beam inside the resin. A computer controlled piezo stage (PI-nanopositioner E-725) allowed the fabrication of structures at various scan speeds. The schematic and actual setup of the two photon lithography (TPL) used are shown in FigureS1 and FigureS2 respectively. After writing, the unexposed regions were washed off using dimethylformamide (DMF) prior to SEM imaging. No post processing of the samples was performed for any measurement.

3. Results and Discussion

Two sets of experiments were performed at a fixed scanning speed of 25 μm/s. The first set of experiments were performed to determine the minimum power required (threshold) for writing at the different LTPO-L concentrations in the polymer. The second set of experiments were performed at a fixed power of 22 mW, which is also the threshold for 1% LTPO-L loading and all the different loading films could be written at this power as determined from first set of experiment. These samples were structurally characterized for their line widths using scanning electron microscopy. The sample micrograph of lines written at threshold powers for 1%, 3%, 5%, 10% and 20% LTPO-L loadings are shown in *Fig2*. A decrease in line width was observed with steady rise in photoinitiator concentration up to 10% of LTPO-L. Sub-wavelength feature sizes of ~ 300nm could be achieved at this loading at laser power ~12 mW (Figure 2 (d)). The size of the polymerized structures can be explained by the growth of voxels, a radical diffusion dominated process, where voxels gradually grow at their active dangling ends as they come in contact with new monomeric units. It is well established that a combination of low laser power and appropriate exposure time are beneficial for the formation of low-aspect ratio voxels [12].

*corresponding author - Tel. no. +91-22-25767615. Email-sshukla@iitb.ac.in

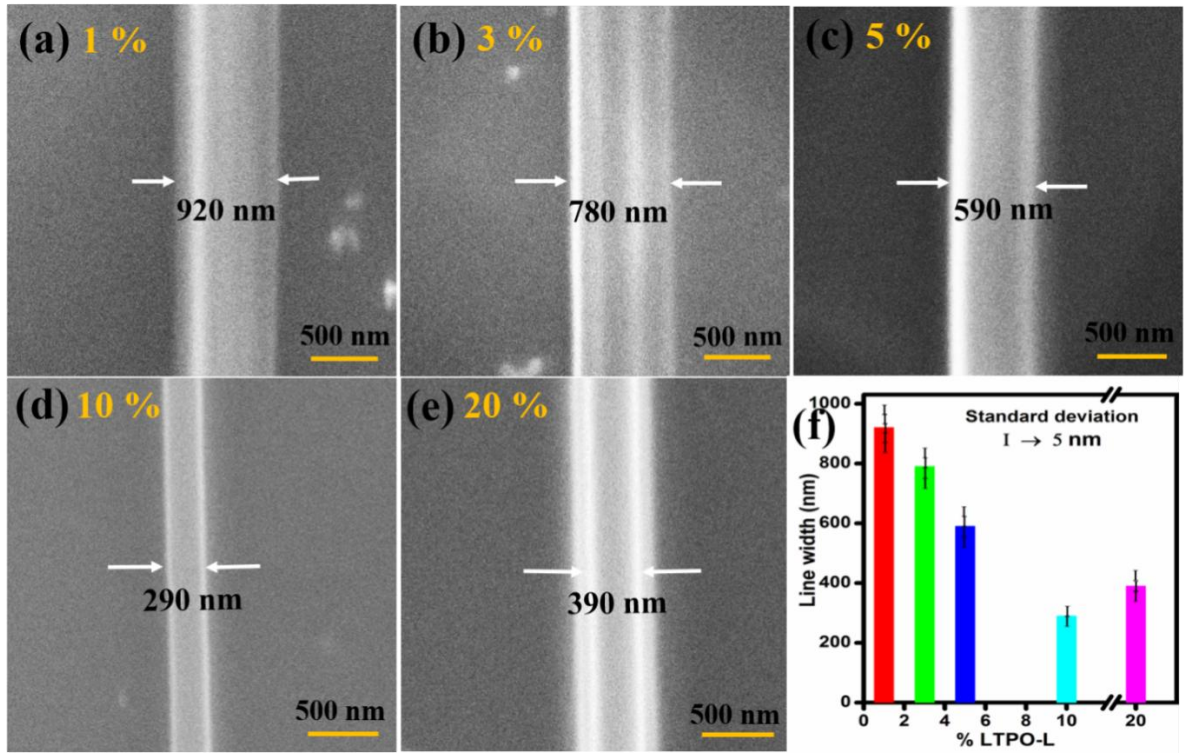

*Fig2: SEM images of lines written using femtosecond laser in a polymeric resin at constant speed of 25 µm/s. (a- e) show lines with increasing photoinitiator concentration from 1%-20% respectively written at threshold powers. (f) Trend of linewidth variation.*

The photoinitiator molecules in the resin absorb incident laser pulses and decompose into radicals through homolytic dissociation. As soon as the radicals are generated they combine with the monomer molecules to initiate the chain polymerization reaction. Solid polymerized volumes are formed only when the monomer conversion ratio is high enough to exceed the gelation point. Even a small number of radicals could trigger photopolymerization provided a higher photon flux is supplied, as noticed in the case of 1% LTPO-L loading. In our experiments the scanning speed (and hence the exposure time) was fixed and LTPO-L concentrations were varied. Hence, the observed trends in the line widths could be attributed to the significant drop in the threshold writing power with increasing photoinitiator loading from 1-10%. As the LTPO-L concentration is further increased above this value to 20% a notable increase in line width was seen. This observation might be explained by the termination mechanism of the chain polymerization reaction. In general the termination of a growing polymeric chain happens by either combining with an active radical or inhibitor. The mechanism of radical quenching can be safely discarded as we have not used any inhibitor molecules. The radical species combine with the reactive center and undergo electron transfer that converts the growing chain into dead polymer which has no further role to play in the polymerization reaction[28]. It is likely that at higher LTPO-L concentrations of 20%, the number of free radicals generated is so high that the rate of polymerization exceeds that of termination which results in escalated linewidths.

*corresponding author - Tel. no. +91-22-25767615. Email-sshukla@iitb.ac.in

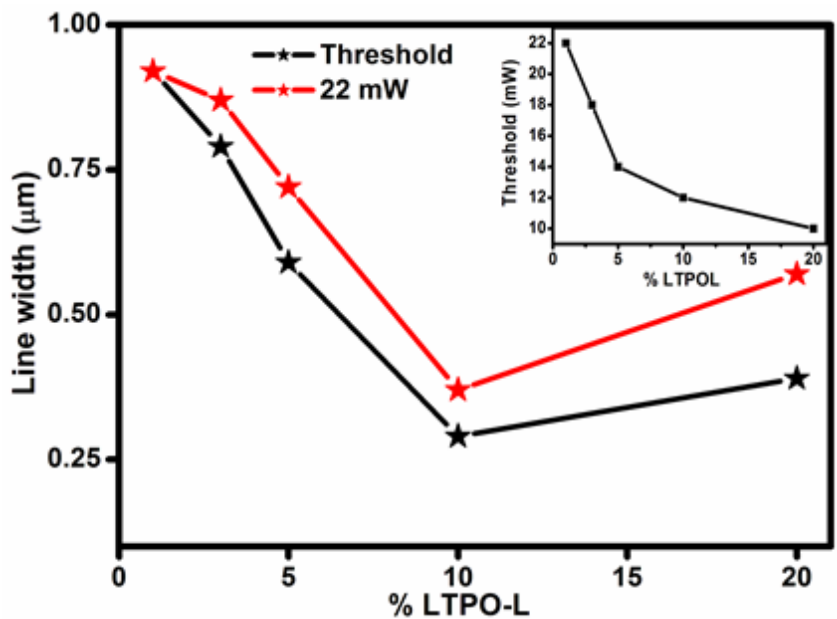

Fig3: Comparison of line widths at threshold laser power and 22mW power at constant scan speed of 25 μm/s for varying LTPO-L concentration. Inset shows the variation of threshold power with increase in LTPO-L concentration.

It can be observed from *Fig3* that the threshold power for writing decreases with increase in photoinitiator concentration. This trend is found to be in good agreement with the literature. It was also observed that lines written at a constant power of 22 mW and speed of 25 μm/s(threshold for 1% LTPO-L loading) showed broader line widths. This could be attributed to the increased fluence beyond the threshold value required for polymerization during writing. The reduction in laser threshold with increasing LPTO-L concentration, as seen in inset in (Figure 3), can be understood as the increase in the number of LTPO-L molecules undergoing cleavage, thereby increasing the production of large number of free radicals that initiate rapid cross linking of the monomeric units.

*corresponding author - Tel. no. +91-22-25767615. Email-sshukla@iitb.ac.in

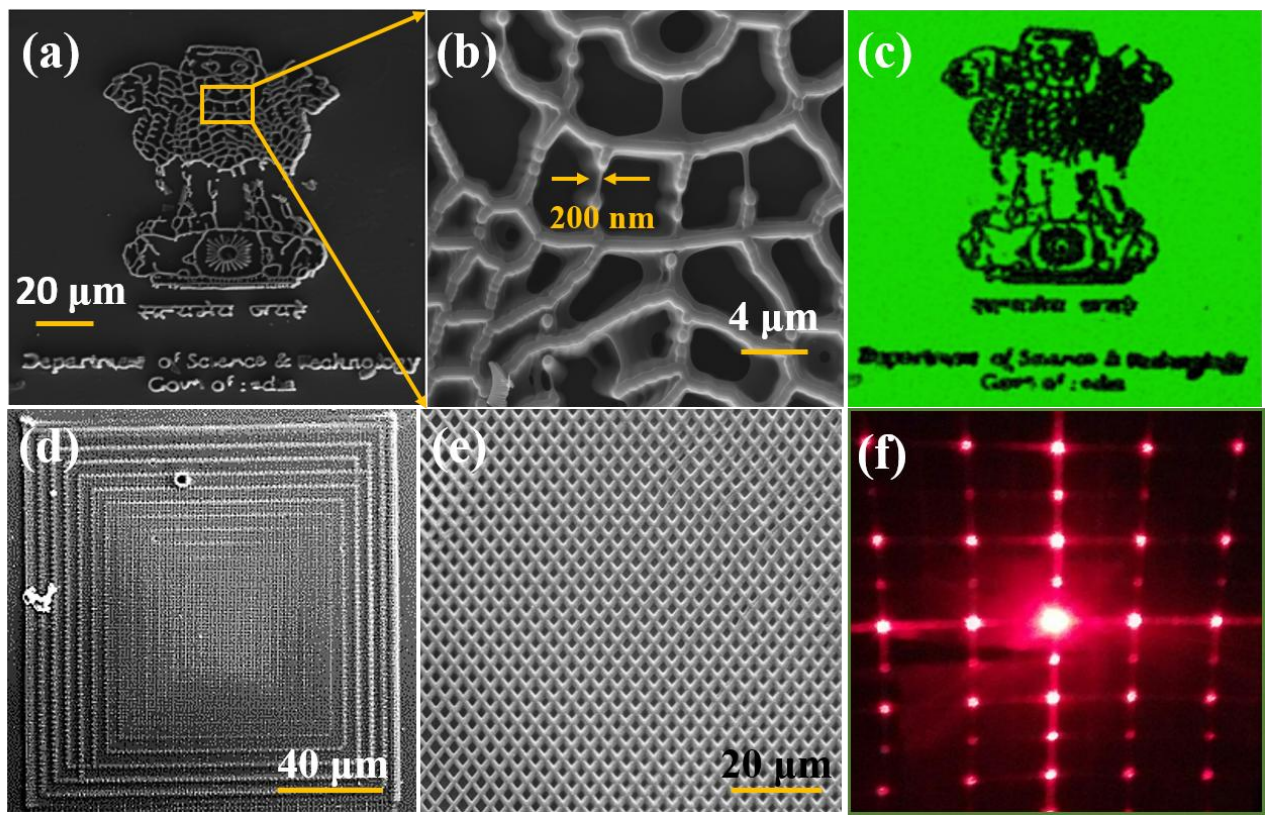

*Fig4:(a) SEM image of DST logo sized 280 µm X 280 µm (b) zoomed in image of the logo showing minimum feature size of 200 nm (c) Optical image of the same logo (d) SEM image of Pyramid (e) Multilayered mesh structure (f) Diffraction pattern of the mesh structure.*

Complex 2D/3D micro/nano structures of various patterns have been written with 10% LTPO-L loading (Figure S3-S7). *Fig4*(a) shows sample SEM micrographs of highly complex 2D structure in form of National Emblem. The dimension of this logo is 280 µm x 280 µm and is written in pure resin at a scanning speed of 1500 µm/s using laser power at 0.20 W at 800 nm wavelength. Figure 4(b) is the zoomed image of 4(a) showing connected structures of submicron and nano feature sizes. (Figure 4(d) and 4(e)) show a pyramid and multilayered mesh structure respectively. The diffraction pattern of the multilayered mesh structure showing bright and intense maxima can be seen in Figure 4(e).

*corresponding author - Tel. no. +91-22-25767615. Email-sshukla@iitb.ac.in

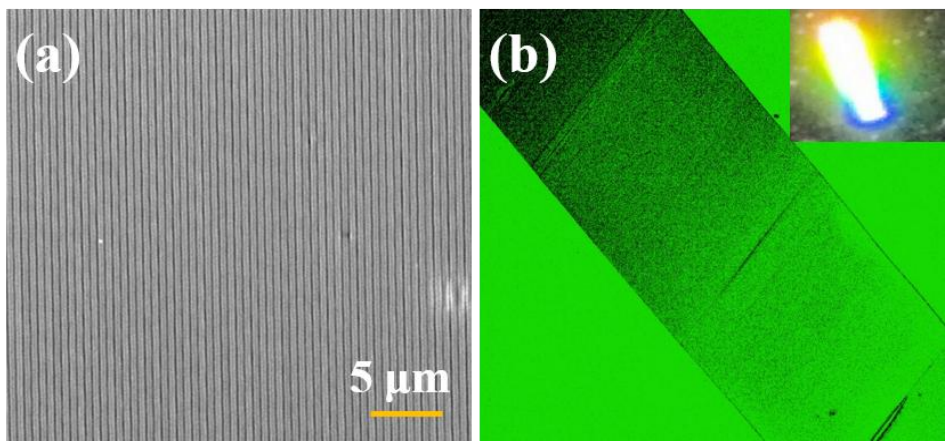

*Fig5: (a) SEM image of Spectral grating with large number of lines/mm, written at a scan speed of 50 µm/s and 150 mW power using 10% LTPO-L. (b) Optical image of the grating. Inset: Optical response of the grating showing different colors in reflectance mode.*

Spectral gratings of variable pitch at scanning speeds of 50 µm/s and 150 mW power were also fabricated by making use of 10% loading concentration of LTPO-L. *Fig5*(a)shows the sample SEM micrograph of a grating consisting of very closely placed thin lines. The optical image of the large area grating (1.12 mm x 280 µm) is shown in Figure. 5(b). These gratings were able to separate the colors (visible to naked eye) from light illuminated from a mobile phone's light source as shown in the inset in Figure 5(b). The low feature sizes obtained using higher loading of photoinitiator enables for the fabrication of large number of lines/mm resulting in very sharp and narrow intensity maximum thus providing high resolutions for spectroscopic applications.

## 4. Conclusions

To summarize, we have demonstrated that critical concentration ofsingle photon photoinitiator can replace the usage of proprietary & expensive two-photon dyes forachieving localized polymerization. This was demonstrated by fabrication of various complex 2D and 3D structures with sub-wavelength resolution. Systematic studies have shown that the line width of the fabricated structures critically depend on and can be controlled by varying the loading concentration of LPTO-L. This can be understood in terms of tradeoff between generation and termination of free radicals for polymerization. Extensive characterization has revealed that a 10 weight % loading of LTPO-L is found to be optimal for patterning sub-wavelength structures. It is concluded that the right balance of power, scan speed and photoinitiator concentration are extremely essential for the fabrication of high resolution structures.

[*]corresponding author - Tel. no. +91-22-25767615. Email-sshukla@iitb.ac.in

**Acknowledgements**  ((This work was supported by the Department of Science and Technology, Solar Energy Research Initiative (SERI), Government of India grant via sanction order no. DST/TM/SERI/2k10/12/ (G).))

[*]corresponding author - Tel. no. +91-22-25767615. Email-sshukla@iitb.ac.in

*corresponding author - Tel. no. +91-22-25767615. Email-sshukla@iitb.ac.in

*corresponding author - Tel. no. +91-22-25767615. Email-sshukla@iitb.ac.in